\documentclass[12pt,preprint]{aastex}

\def\princeton{1}
\def\stanford{2}
\def\riken{3}
\def\cfa{4}
\def\telaviv{5}

\shorttitle{Strongly lensed E+A galaxy}
\shortauthors{Shin et al.}

\begin{document}

\title{THE SDSS DISCOVERY OF A STRONGLY 
LENSED POST-STARBURST GALAXY AT z=0.766}
\author{Min-Su~Shin\altaffilmark{\princeton},
Michael~A.~Strauss\altaffilmark{\princeton}, 
Masamune~Oguri\altaffilmark{\stanford},
Naohisa~Inada\altaffilmark{\riken}, 
Emilio~E.~Falco\altaffilmark{\cfa}, 
Tom~Broadhurst\altaffilmark{\telaviv}, 
James~E.~Gunn\altaffilmark{\princeton}}

\altaffiltext{\princeton}{Princeton University Observatory,
Peyton Hall, Ivy Lane, Princeton, NJ 08544}
\altaffiltext{\stanford}{Kavli Institute for Particle Astrophysics and Cosmology, Stanford University, 
2575 Sand Hill Road, Menlo Park, CA 94025}
\altaffiltext{\riken}{Cosmic Radiation Laboratory, RIKEN (The Physical and Chemical Research 
Organization), 2-1 Hirosawa, Wako, Saitama 351-0198, Japan}
\altaffiltext{\cfa}{Harvard-Smithsonian Center for Astrophysics, 60 Garden Street, 
Cambridge, MA 02138}
\altaffiltext{\telaviv}{School of Physics and Astronomy, Tel Aviv University, Tel Aviv 
69978, Israel.}

\begin{abstract}

 We present the first result of a survey for strong galaxy-galaxy lenses in Sloan 
Digital Sky Survey (SDSS) images. 
SDSS J082728.70+223256.4 was selected as a lensing candidate using selection 
criteria based on the color and positions of 
objects in the SDSS photometric catalog. 
Follow-up imaging and spectroscopy showed this object to be a lensing system. 
The lensing galaxy is 
an elliptical at z = 0.349 in a galaxy cluster. The lensed galaxy has the spectrum of a 
post-starburst galaxy at z = 0.766. The lensing galaxy has an estimated mass of 
$\sim$ 1.2$\times$ $10^{12}\ {\rm M_{\odot}}$ and the corresponding mass to light 
ratio in the $B$-band is $\sim$ 
26 ${\rm M_{\odot}/L_{\odot}}$ inside 1.1 effective radii of the lensing galaxy. Our study shows how 
catalogs drawn from multi-band surveys can be used to find strong galaxy-galaxy 
lenses having multiple lens images. Our strong lensing candidate selection based on photometry-only catalogs will be useful 
in future multi-band imaging surveys such as SNAP and LSST.

\end{abstract}


\setcounter{footnote}{0}

\section{INTRODUCTION}

Gravitational lensing is a highly useful phenomenon in astrophysics. It allows us to 
study faint galaxies that have been magnified, and can be used for a variety of 
cosmological constraints. For example, a 
strong lens forming multiple images of a background source allows us to determine 
the mass distribution of the lensing object \citep[e.g.][]{bolton06}. 
Moreover, the discovery of a substantial number of lens systems makes it possible 
to study the potential wells of lensing galaxies in a statistical way \citep[e.g.][]
{koopmans06}. The lensed galaxy itself is magnified, allowing us to study its 
properties in detail \citep[e.g.][]{seitz98,richard06}. When a background 
quasar is strongly lensed, this magnification sometimes makes the host galaxy 
visible \citep{peng06}. The statistics of strong lenses allow constraints on the 
cosmological parameters \citep[e.g.][]{kochanek03,oguri07a}, including models of dark energy \citep{fedeli07,oguri08}.

Because of the importance of gravitational lensing systems, there have been many systematic studies of 
them and surveys for them. 
In particular, strong gravitational lensing of quasars has been studied systematically by the CASTLES 
collaboration \citep{castles98}, the CLASS \citep{class01},  
and the Sloan Digital Sky Survey (SDSS) \citep{inada07}. 
While various surveys of galaxy lensing have been carried out 
\citep[e.g.,][]{fassnacht04,moustakas07}, many discoveries of lensed galaxies 
have been serendipitous, such as the discovery of the lensed galaxy cB58 at z=2.7 
\citep{yee96,seitz98}. 
Systematic searches for strongly lensed galaxies 
have been also made in the fields of massive clusters \citep[e.g.][]{pello05}. 
For example, HST WFPC2 images were used to find 
giant arcs around galaxy clusters \citep{sand05}. \citet{hennawi08} obtained 
deep images of 240 rich galaxy clusters selected from SDSS, discovering 16 new giant arc systems. 
These giant galaxy cluster lenses allow us to determine the mass distribution on scales of several hundred kpc.

The large volume spectroscopic survey of the SDSS \citep{york00} allows systematic searches  
for galaxy-galaxy lenses. \citet{bolton04} and \citet{willis06} searched for objects whose 
spectra show features of both low-redshift ellipticals 
and higher-redshift emission lines. 
Deep HST imaging confirms that the foreground elliptical is lensing the background galaxy. 
This method has discovered enough strong lensing cases to allow a statistical study of 
lensing galaxies \citep[e.g.][]{bolton06,bolton07}.

Using SDSS imaging data, \citet{allam07} serendipitously discovered a lensed Lyman break galaxy at z = 
2.73 around a lensing galaxy at z = 0.38. The lensing galaxy of this lens system has an SDSS 
spectrum, which does not show any features of the lensed 
background galaxy because the separation between the lensing galaxy and the lensed image is larger 
than the 1\farcs5 radius of the spectrograph fiber. 
This lens therefore did not enter the \citet{bolton04} sample.

These discoveries suggest an automated search for galaxy lenses in wide-field imaging data. 
Several recent studies have used automatic image analysis techniques to find strong lenses 
in deep images \citep{lenzen04,horesh05,alard06,
seidel06,cabanac07}. \citet{estrada07} have also applied a similar method to SDSS images to 
find giant arcs around clusters of galaxies. These methods use the characteristic arc shape of 
lens images by calculating the ellipticity or the length-to-width ratios of objects in the imaging 
data.  These arc-shape detection programs can find a hidden 
compact ring around a lensing galaxy, as shown by \citet{cabanac07}.

However, strong lensing detection based on the shape of objects has several difficulties. 
First, this kind of survey depends on several parameters that are tuned 
for an optimal detection rate. For example, when the properties of the images such as signal-to-noise 
ratio and seeing change, the parameters have to 
be retuned before the codes are applied to the new images. Second, it does not work 
well when the images are not deep enough to show a bright arc shape, 
even if the lensed galaxy is quite visible. Third, it is computationally demanding 
to process massive quantities of wide field image data with shape detection codes 
when we want to search all lensing possibilities around all elliptical galaxies. 
A simpler method to search for strong galaxy-galaxy lensing would be valuable.

We find that a lensing search based only on a photometric catalog can be used to detect strong galaxy-galaxy 
lenses in SDSS images, even though they are not deep enough to show the full morphology 
of lensed arcs. 
All SDSS imaging data are processed by an automated pipeline \citep{lupton02}. Measured 
parameters of detected objects including multi-band 
magnitudes and correlations with neighboring objects are archived 
to a database from the pipeline. 
Indeed, the photometric catalog has enough information to allow us to 
separate the rare lenses from other objects.

In this paper we show the first results of our search for strong galaxy-galaxy lenses in SDSS. 
We explain the candidate selection strategy in the following section. 
In \S3, we present follow-up observational data of the first confirmed lensing object,  
SDSS J082728.70+223256.4. A possible lens model is given in \S4. 
The properties of the lensing galaxy are described in \S5. 
Discussion and conclusions follow in \S 6. 
In this paper we focus on demonstrating the use of a 
multi-band photometry catalog to find lensing systems. 
We are still refining our selection algorithm, so we will quantify the completeness 
of our sample and give details of our final candidate selection method in 
a subsequent paper.

\section{CANDIDATE SELECTION}

The SDSS produces a photometric catalog in five bands using a 2.5m telescope equipped with 30 
CCDs \citep{gunn98,gunn06} as well as a spectroscopic catalog of objects selected from 
the imaging data. The properties of all objects detected above 5 $\sigma$ in at least one band 
over the ${\rm \sim\ 10^{4}\ deg^{2}}$ imaged to date are measured by a dedicated pipeline \citep{lupton02} and 
are calibrated photometrically \citep{tucker06} and astrometrically \citep{pier03}. 
We use the fifth public data release \citep{dr5} in selecting lens candidates. Since most possible lensing galaxies are early-type galaxies 
because of their large velocity dispersions \citep{turner84, fukugita91}, we assume that lensing galaxies are red galaxies in the SDSS images. 
Because only the most luminous of those possible lensing galaxies have redshifts 
above 0.2 \citep{strauss02,eisenstein01}, 
we do not restrict our analysis to the sample of objects with redshifts.

Our candidate selection method is still being developed, so we only outline the criteria that were used 
in finding the reported lens system here. 
First, we identify all red galaxies in the photometric catalog that have a model magnitude in 
$g$-band $<$ 23.0 without correction for Galactic extinction. Red galaxies are defined using color cuts suggested 
by \citet{goto02}:
\begin{eqnarray}
-0.2 < (r - i) - 0.25 (g - r) - 0.18 < 0.2, \\
-0.2 < (g - r) < 2.0,
\end{eqnarray}
where we use model magnitudes throughout. This color cut selects intrinsically red galaxies up to redshift 0.4. 
We next select those red galaxies that have at least 
two photometric 
companions between 2\farcs4 and 3\farcs6 away, a typical radius of strongly 
lensed systems, and a large enough separation that the image deblender separates the lens images 
from the red galaxy.

In order to discriminate a lensing system from a galaxy group or cluster, we require that the $u - g$ colors of 
the possible lens images differ by more than two magnitudes from that of the lensing galaxy. 
This color difference is much larger than the observed color range of cluster ellipticals \citep[e.g.][]
{goto02}. This color difference cut was adopted to test our preliminary selection method and to reduce possible 
contamination of neighbor galaxies around red galaxies. While cluster ellipticals are red and tend to 
be undetected in $u$, background lensed galaxies are often quite blue and have very different $u - g$ colors. 
We discuss refinement of this color cut in \S5.

In most strong quadruple lensing systems we expect to find at least two bright knots on one side of the lensing 
galaxy \footnote{The case of double images is discussed in \S5.}. We therefore require that the angle between the two 
companion objects with respect to the possible lensing galaxy be less than 90$^{\circ}$. Even though the surface 
brightness of the arc may not be high enough to appear in the SDSS images, the positions of the bright knots 
are precise enough to check whether the positions of the photometric objects are consistent with a 
lensing system.

Finally, we eye-inspect the objects that satisfy our criteria. We found 38154 galaxies that satisfy 
the color cut and have at least one companion with $u - g$ color difference of more than two magnitudes 
in the separation bin of 2\farcs4 and 3\farcs6. 
Among the galaxies found, 1644 objects have more than one companion. But 
only 38 objects have an angle of less than 90$^{\circ}$ between the two lens images; these 
we examine by eye.

From its appearance in the SDSS images (Figure 1), SDSS J082728.83+223253.9 (G) was the most promising 
lensing galaxy in our search. 
The basic SDSS photometry data are  
given in Table 1. The model magnitude $u - g$ color of this elliptical galaxy is 4.48 without 
Galactic extinction correction. The lensing galaxy 
is undetected in $u$, as expected for a red elliptical. Therefore, the $u - g$ color 
of the lensing 
galaxy should be interpreted as a lower limit. Meanwhile, the 
lens images are deblended as three photometric objects: SDSS J082728.87+223256.5 
(A), SDSS J082728.70+223256.4 (B), and SDSS J082728.86+223252.6 (C). 
The first two objects are classified as galaxies by the SDSS software 
while the last one is star-like. The distances of objects A and B from 
the lensing galaxy are $\sim$ 3\farcs0, and satisfy our selection criteria. 
Although image C is also close to the lensing galaxy, its model magnitude in 
the $g$-band is fainter than 
23 mag, and therefore was not selected by our criteria. The angle between images 
A and B is $\sim$ 40\degr, i.e. less than 90\degr.   Images A and B have 
$u - g$ = 0.36 $\pm$ 0.65 and $-0.01 \pm 0.08$ without Galactic extinction correction, 
respectively. The larger color error for A simply reflects the fact that 
it is 2 magnitudes fainter than B. The difference of $u - g$ between the lensing 
galaxy and the lens images is larger than the color error itself. 
The fact that two objects with different $u - g$ colors from the 
lensing galaxy are on the same side of the galaxy strongly suggests that 
they are images of a gravitationally lensed background galaxy.

\section{FOLLOW-UP OBSERVATIONS}

\subsection{APO 3.5m Spectroscopy}

We observed the brightest object A using the Double Imaging Spectrograph (DIS) of the Astrophysical Research 
Consortium 3.5m telescope at Apache Point Observatory (APO) on 2006 December 17 (UT). We set the 
two gratings to cover the wavelength ranges from $\sim 3700 {\rm \AA}$  to $\sim 5570 {\rm \AA}$  and from 
$\sim 5300 {\rm \AA}$  to $\sim 9000 {\rm \AA}$, with dispersions of 
1.83 ${\rm \AA}$ per pixel and 2.31 ${\rm \AA}$ per pixel for the blue and red channels of DIS, respectively.
We exposed for 2400 seconds using a 1\farcs5 slit. The spectrum given in Figure 2 has signal-to-noise 
ratio of $\sim$ 7.5 per pixel 
around $\lambda \sim 6500 {\rm \AA}$, allowing us to measure the redshift and to identify the 
characteristics of the object. The spectrum shows strong Balmer absorption lines of 
redshift z=0.7654 $\pm$ 0.0004, characteristic of post-starburst 
galaxies \citep{goto03}. The spectrum does not show any strong emission lines. 
E+A galaxies are 
rare objects, representing $1 - 2$\% of the galaxy population at z$\approx$0 \citep{quintero04,yan06}, 
and lensed E+A galaxies are rarer still; there is only one previously known lensed E+A galaxy, at 
z = 1.394 \citep{fassnacht96}. 

The spectrum has a blue continuum and shows 
strong absorption by MgII $\lambda\lambda 2796, 2803$, 
although the doublet is not resolved in our low-resolution spectrum. As shown 
in \citet{tremonti07}, the MgII absorption might be a feature of an outflow in 
this post-starburst galaxy.  Because outflows in post-starburst galaxies are 
thought to originate from galactic winds, the spatial location of 
this absorption feature in the galaxy would be valuable for understanding 
the galactic wind scenario. 
Thanks to the strong lensing of this galaxy, 
spatially resolved spectroscopy should allow us to map the MgII absorption feature across the galaxy.

\subsection{Subaru Imaging and Spectroscopy}

We conducted imaging and spectroscopic follow-up observations with the
Faint Object Camera And Spectrograph \citep[FOCAS;][]{kashikawa02} on the
Subaru 8.2-meter telescope on 2007 January 21. The spectroscopic
observation was performed with 2x2 on-chip binning with the 
300B grism and SY47 filter. A 1\arcsec\ slit was aligned 
to observe objects B and C and  the lensing galaxy G 
simultaneously. An 1800 sec exposure was taken with seeing of $\sim 0\farcs7$. 
We also took 120 sec $V$- and $I$-band images. All the data were reduced using 
standard tasks in IRAF.

The spectra of the lens images B and C in Figure 2 show the same redshift as object 
A. The redshift of object B, z=0.7659 $\pm$ 0.0003, is 
determined by the absorption lines H8, H7, H6, H, and H$\delta$. The 
lines are also used in estimating the redshift of object C, resulting 
in z=0.7660 $\pm$ 0.0007.

The spectrum of the lensing galaxy G (Figure 2) has a redshift of 
z= 0.3492 $\pm$ 0.0007 based on the CaII K \& H, G-band, and Mg lines. 
This is in good agreement with the SDSS DR5 photometric redshift 
\citep{csabai03}, 0.305 $\pm$ 0.039.  An object 13\farcs2 away, 
SDSS J082728.62+223241.0, has an SDSS redshift of z=0.3347 $\pm$ 0.0002. 
The lensing galaxy is surrounded by other red galaxies of similar color, 
suggesting a similar redshift.  The lensing galaxy thus appears to be a member of a 
galaxy cluster. The case for this is strengthened by the 
presence of a X-ray source in the ROSAT All-Sky Survey Faint 
Source Catalog \citep{voges00}, 1RXS J082726.8+223237, that is $\sim 33$'', i.e. 
a projected distance of $\sim 160$ kpc, 
away from the lensing galaxy. The positional error of this faint 
source is $\sim 14$'' \citep{voges00}, thus this position is consistent with that 
of the apparent cluster of galaxies. 
Assuming that the X-ray source is at the same redshift as the lensing galaxy, 
the estimated flux from the ROSAT catalog is about $6\times10^{-13}~{\rm erg/cm^{2}~s}$ with 
the Galactic absorption correction, assuming a Raymond-Smith plasma spectrum 
\citep{raymond77} or a simple power-law spectrum of index -0.5. 
It corresponds to a luminosity of about $2\times10^{44}~{\rm erg/s}$ 
in the range 0.5 - 2.0 keV. This value is close to the luminosities of typical 
galaxy clusters \citep{mullis03}. The confirmation of the 
galaxy cluster is discussed further in \S5.

The Subaru $V$-band image (Figure 3) shows the blue lensed objects more distinctly 
than SDSS, because it is deeper and has better seeing. 
We measure the positions of the brightness peaks using SExtractor \citep{bertin96}. To deblend the 
faintest lens image C from the lensing galaxy, we used a Mexican hat filter with a full-width at 
half-maximum of 0\farcs6. The derived relative positions 
of the lens images are used to constrain a lens model in \S4.

We suspect that the faint object D, at 082729.11+223253.76 in Figure 3 is 
the fourth lens image. 
In order to check this hypothesis, we measure $(V - I)$ color differences among the lens 
images. The color is measured through a 3\arcsec\ diameter aperture. Between objects D 
and A, $\Delta(V - I) = (V - I)_{D} - (V - I)_{A}$ is 0.13 $\pm$ 0.04 while $\Delta(V - I)$  
between D and B is 0.20 $\pm$ 0.04, where the errors are from photon statistics only. 
Given the uncertainties in deblending, this agreement is 
consistent with the hypothesis that D is the possible fourth lens image.

\section{THE LENS MODEL}

We fit a model of a singular isothermal ellipsoid (SIE) with external shear to 
the positions of the four components A, B, C and D 
system using the {\it gravlens} software \citep{keeton01}; this is the simplest 
model which can fit quadruple lenses in general \citep{keeton97}. 
However, the cluster environment of the lens (\S5) suggest a more complicated 
model \citep[e.g.][]{keeton04,kawano04}. The perturbation of mass substructures 
in the lensing environment will affect the lensing magnification \citep[e.g.][]{saha07}. 
For this reason, we do not use relative flux ratios among the lens images in 
constraining the lens model, but just use their relative positions.

A well-fitted lens model is found with a positional error of $\sim$ 0\farcs2 between 
the derived and observed positions of lens images, as shown in Figure 4. 
The derived central position of 
the lensing galaxy is within 0\farcs1 of the optical position. 
The required external shear is about 0.1 in this model. 
The direction of the critical curve is completely inconsistent with the position angle 
of the lensing galaxy. But this discrepancy could be explained by 
the dominance of the external shear. 
The position angle of the quadrupole term is within $5^{\circ}$ of the direction 
to the brightest cluster galaxy. This angle difference is acceptable when 
considering the possible offset between the position of the brightest cluster 
galaxy and the center of mass of the cluster, and the degeneracy between 
the strength and position angle of the external shear. The Einstein radius (${\rm R_{Ein}}$) 
from the fitted model is 2\farcs23$\pm$0.01\arcsec, corresponding to 10.8 kpc in the lens plane 
\footnote{We adopt cosmological parameters of $\Omega_{{\rm m}}=0.26$, 
$\Omega_{\Lambda}=0.74$, and h=0.72 as given in \citet{spergel07}. The error of 
the ${\rm R_{Ein}}$ represents only a fitting error of the mass parameter.}. 
${\rm R_{Ein}}$ is the mass parameter of a singular isothermal ellipsoid, that is related 
to the velocity dispersion of the lensing galaxy as used in \citet{keeton04}. 
The configuration of the source and lens images is consistent with what we expect from 
strong lensing in a galaxy cluster environment \citep[e.g.][]{king07}.

\section{THE LENSING GALAXY}

We present a simple estimation of the properties of the lensing galaxy. A detailed lensing model 
and other properties of lensing galaxies will be given as part of a study of a larger sample. 

We can estimate a rough total mass of the lensing galaxy inside ${\rm R_{Ein}}$ derived in the 
previous section:
\begin{equation}
M = \frac{R_{Ein}^{2} c^{2}}{4 G} \frac{D_{d} D_{s}}{D_{ds}} \sim 1.2 \times {\rm 10^{12}\ M_{\odot}},
\end{equation}
where $D_{s}$, $D_{d}$, and $D_{ds}$ are angular-diameter distances to the 
lensed galaxy, to the lensing galaxy, and between the two galaxies, respectively. 
The SDSS photometric pipeline gives a de Vaucouleurs effective radius of the lensing 
galaxy of 1\farcs96 
$\pm$ 0.17\arcsec, or 9.5 $\pm$ 0.8 kpc in the SDSS $i$-band. Therefore, ${\rm R_{Ein}}$ is roughly 
1.1 effective radii. The velocity dispersion of the lensing galaxy is poorly measured at 210 
$\pm$ 130 km/s due to the low resolution of the spectrum, R = 500. With this large uncertainty, 
it is not meaningful to compare this velocity dispersion with that derived from the SIE model.

We next estimate a mass-to-light ratio. The model magnitude is 21.0 in 
the $g$-band within 
1.1 effective radii, 
after correction for Galactic extinction from \citet{schlegel98}. 
Therefore, the absolute magnitude of the lensing galaxy is $-21.2$ 
in the $B$-band, applying the K-corrections given in \citet{fukugita95}. 
The estimated ${\rm M/L}$ of the lensing galaxy is 
$\sim 26\ {\rm M_{\odot}/L_{\odot}}$ in the $B$-band. 
We have taken the absolute magnitude of the Sun to be 5.45 mag 
in the $B$-band \citep{blanton07}.

This mass-to-light ratio is higher than 
that of typical elliptical galaxies, 
$\sim$ 10 ${\rm M_{\odot}/L_{\odot}}$ 
\citep{bahcall95,keeton98,wilson01,padmanabhan04}, and is 
comparable to that of a 
galaxy group \citep{bahcall95,parker05}. 
This mass-to-light ratio is acceptable when we remember 
that the lensing galaxy is near the 
center of a galaxy cluster as shown in \S3.2, and the lensing 
is sensitive to all the projected mass. Most massive early-type 
galaxies reside in dense regions like galaxy clusters \citep[e.g.,][]{kauffmann04}. 
Indeed, many lensing ellipticals are known to lie in 
galaxy group or cluster environments \citep{fassnacht02,williams06,auger07}, 
partly because of enhancement of the strong lensing probability by 
external convergence from dark matter in the group/cluster \citep{oguri05}. 
Note that the redshift difference between the lensing galaxy and 
the SDSS spectroscopic object SDSS J082726.8+223241.0 is equivalent 
to a radial velocity difference of 
$\sim$ 3000 km/s. This implies either that this galaxy cluster is quite massive, or 
that there are multiple superposed clusters along the line of sight. Further 
spectroscopy of galaxies in the field is needed to distinguish these cases.

Figure 5 shows the color-magnitude diagram and $(g - r)$ color distribution of 
galaxies within 3\arcmin, i.e. $\sim$ 870 kpc on the lens plane, around the lensing 
galaxy. The lensing galaxy appears to be a member of a galaxy cluster, although 
it is not the brightest member. As shown 
in \citet{goto02}, the expected $(g - r)$ color of ellipticals 
at z $\sim$ 0.3 is $\sim$ 1.6. We select ellipticals with $r$-band model magnitude $<$ 21.5 using 
the criterion that the de Vaucouleurs model fitting likelihood is 10\% larger than that of 
the exponential model. There is a clear concentration of ellipticals around the expected 
$(g - r) \sim$ 1.6. The brightest member galaxy SDSS J082728.42+223245.7 is about 10\arcsec\ away 
from the lensing galaxy. That is, the projected distance is about 48 kpc. Therefore, the 
lensing might be affected by the dense central part of the galaxy cluster, 
as we argued above.

\section{DISCUSSION \& CONCLUSION}

We have shown that it is possible to detect strong galaxy-galaxy lenses 
with photometric catalogs drawn from large multi-band imaging surveys. 
Thus our method can be applied to the next-generation of 
wide-field multi-band surveys including UKIDSS \citep{ukidss}, 
Pan-STARRS \citep{kaiser04}, SNAP \citep{kim02}, 
LSST \citep{tyson05}, the Sky Mapper Southern Sky Survey \citep{keller07}, and VISTA \citep{vista}.

In SDSS and future surveys, our approach can be improved 
using photometric redshifts as well as color information; 
the precision of photometric redshifts is good enough to discriminate low-redshift elliptical galaxies from 
the high-redshift lensed galaxy. Moreover, photometric-redshift techniques also allow an estimate of 
the intrinsic colors of galaxies \citep{csabai03}, allowing red galaxies to be identified in a more sophisticated 
way than we have.

The photometric pipelines and databases of future multi-band wide-field imaging surveys must have several 
components in order for our approach to work. 
First, the detection of the lens system described in this paper was possible due to the 
successful deblending algorithm used by the SDSS photometric pipeline \footnote{\url{http://www.sdss.org/dr5/algorithms/deblend.html}} \citep{lupton02}. When the lensing arc is not bright enough to be classified as an 
extended object, the lens images often look like a series of disconnected, deblended knots, as we saw in Figure 1.
So, we expect our approach to have 
better efficiency for detecting faint lensed galaxies than do image analysis methods. 
The second important consideration is a database which links close neighbors together, 
given our criteria on the angular distance between possible lensing galaxies and lensed images.

Given that our method uses position and color information only, it can easily be used with archival data in the Virtual 
Observatory. 
In particular, the application of our approach to combined catalogs over a range of wavelengths will benefit 
from improved photometric redshifts. For example, the recent discovery of 
high-redshift galaxy clusters with massive dark matter 
halos \citep[e.g.][]{stanford05,brodwin06} suggests applying our method to the database of a deep multi-band 
imaging survey of high-redshift clusters.

We note that our strategy has several disadvantages. First, our method is restricted by the angular resolution of 
the data. The poor angular resolution of SDSS, $\sim$ 1\farcs2, does not permit 
us to find compact lenses. 
Future surveys like LSST will be much superior in that regard. 
Second, our selection criteria are based on our incomplete understanding of the properties of possible lensing 
galaxies. As we mentioned in \S2, we 
narrowed the color range of possible lensing galaxies to increase the lensing 
detection probability. 
Third, our method will be contaminated  
by foreground objects mistaken as possible lens images. Improved photometric 
redshifts should reduce this problem.

We continue to refine our lens candidate selection method. First of 
all, our requirement of a color difference of more than two magnitudes between the candidate lensed 
and lensing galaxies is rather crude, and can be replaced with a criterion based on photometric 
redshifts. In addition, we can use other colors like $(g - r)$ or $(u - r)$ to separate lenses from 
galaxy clusters. In particular, early-type galaxies have a narrow range of $(g - r)$ color, suggesting 
that $(g - r)$ color differences could be better discriminant of background lensed galaxies. 
We can also loosen the criterion on the angular separation between the two lens images 
to find double lens systems, and the criterion on separation between a lensing galaxy and images also can be 
modified to include large-separation lenses.

We have described a new method to search for strong galaxy-galaxy lenses from the SDSS photometric catalog. 
We have demonstrated the use of our method with the discovery of a new strong galaxy lens. 
A lensing elliptical galaxy at z = 0.349 in 
a galaxy cluster magnifies a background post-starburst galaxy at z = 0.766, producing four images. 
Our selection method will be useful for future imaging surveys, with their superior depth and angular 
resolution.

It is possible to combine our lensing search method based on a photometric catalog with 
a direct image analysis method \citep[e.g.][]{cabanac07}. By using both approaches, we will be able to 
detect various classes of lenses, ranging from faint lens images to giant arcs around galaxy 
clusters. For example, 
after selecting a possible lensing system using our method, one can easily apply the image analysis 
method to the selected object, reducing the time required to apply the 
direct image analysis approach to a massive imaging dataset.

We are refining our selection criteria and plan follow-up observations of 
several candidates. 
We will publish our full sample and present details of lens models for each object, 
statistical properties of the lenses, and the detection statistics of our lens sample 
in future papers. The high magnification that the lensing gives allows us to study 
the populations of these otherwise 
faint and distant galaxies in detail. In particular, 
for the lensed E+A galaxy we present here, we plan observations using an 
integral field unit spectrograph and adaptive optics to understand the spatial distribution of the MgII 
absorption and blue continuum; given a precise lensing model, we may be able to spatially resolve the 
outflow across the face of the galaxy.

\acknowledgments

M.-S.~S. and M.~A.~S. acknowledge the support of NSF grants AST-0307409 and 
AST-0707266. 
N.~I. acknowledges support from the Special 
Postdoctoral Researcher Program of RIKEN. 
We thank 
Huan Lin, Sahar Allam, Douglas Tucker, Tom Diehl, and James Annis of the SDSS group 
at the Fermi National Accelerator Laboratory for many fruitful discussions. 
This work was supported in part by Department of Energy contract DE-AC02-76SF00515.

This paper is based in part on observations 
obtained with the Apache Point Observatory 
3.5-meter telescope, which is owned and operated by the Astrophysical 
Research Consortium and based on data collected at Subaru Telescope, which 
is operated by the National Astronomical Observatory of Japan.

Funding for  the SDSS and SDSS-II  has been provided by  the Alfred P.
Sloan Foundation, the Participating Institutions, the National Science
Foundation, the  U.S.  Department of Energy,  the National Aeronautics
and Space Administration, the  Japanese Monbukagakusho, the Max Planck
Society, and  the Higher Education  Funding Council for  England.  The
SDSS Web  Site is  http://www.sdss.org/.  The SDSS  is managed  by the
Astrophysical    Research    Consortium    for    the    Participating
Institutions. The  Participating Institutions are  the American Museum
of  Natural History,  Astrophysical Institute  Potsdam,  University of
Basel,   Cambridge  University,   Case  Western   Reserve  University,
University of Chicago, Drexel  University, Fermilab, the Institute for
Advanced   Study,  the  Japan   Participation  Group,   Johns  Hopkins
University, the  Joint Institute  for Nuclear Astrophysics,  the Kavli
Institute  for   Particle  Astrophysics  and   Cosmology,  the  Korean
Scientist Group, the Chinese  Academy of Sciences (LAMOST), Los Alamos
National  Laboratory, the  Max-Planck-Institute for  Astronomy (MPIA),
the  Max-Planck-Institute  for Astrophysics  (MPA),  New Mexico  State
University,   Ohio  State   University,   University  of   Pittsburgh,
University  of  Portsmouth, Princeton  University,  the United  States
Naval Observatory, and the University of Washington.

%------------- REFERENCE ----------------%

%------------- TABLE ----------------%

\begin{deluxetable}{ccccccc}
\tabletypesize{\small}
\rotate
\tablewidth{0pt}
\tablecaption{SDSS model magnitudes of objects making up the lens system.}
\tablehead{\colhead{object} & \colhead{coordinate} & \colhead{u} & \colhead{g} & \colhead{r} & \colhead{i} & \colhead{z}}
\startdata
A & SDSS J082728.87+223256.5 & 21.93 $\pm$ 0.62 & 21.57 $\pm$ 0.19 & 24.23 $\pm$ 2.29 & 23.39 $\pm$ 1.96 & 21.27 $\pm$ 0.95 \nl
B & SDSS J082728.70+223256.4 & 20.17 $\pm$ 0.07 & 20.18 $\pm$ 0.03 & 20.27 $\pm$ 0.05 & 19.42 $\pm$ 0.03 & 19.38 $\pm$ 0.09 \nl
C & SDSS J082728.86+223252.6 & 22.52 $\pm$ 0.27 & 24.11 $\pm$ 0.40 & 24.80 $\pm$ 0.66 & 23.9 $\pm$ 0.59 & 22.39 $\pm$ 0.43 \nl
G & SDSS J082728.83+223253.9 & 24.93 $\pm$ 2.07 & 20.45 $\pm$ 0.04 & 18.86 $\pm$ 0.02 & 18.24 $\pm$ 0.02 & 17.82 $\pm$ 0.03 \nl
\enddata
\tablecomments{Galactic extinction correction is not applied in this table.}
\end{deluxetable}

%------------- FIGURE ----------------%

\begin{figure}
\plotone{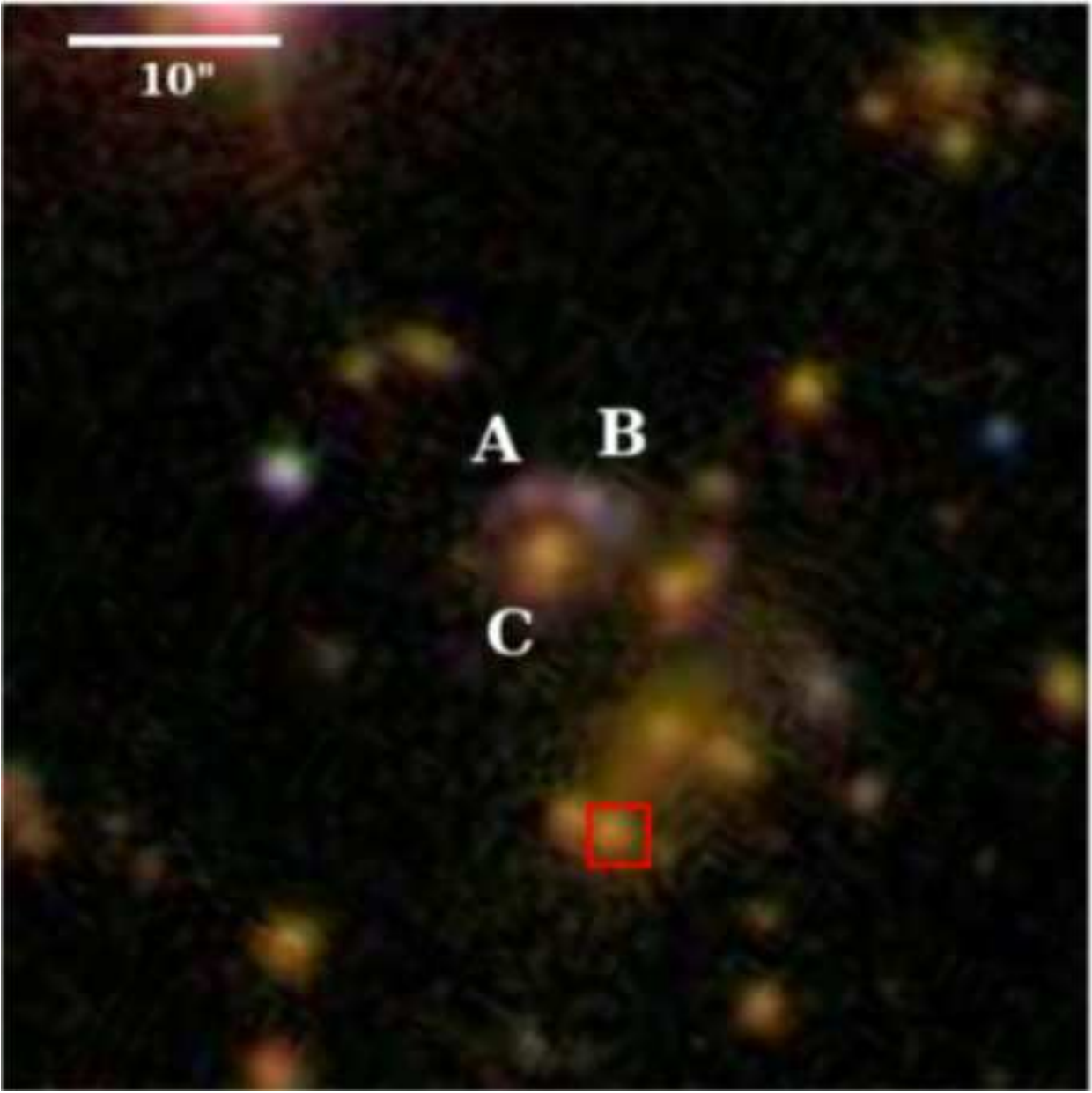}
\caption{SDSS gri composite cutout image of the lensing system, 
with seeing of $\sim$ 1\farcs2. 
Three lens images (A, B, and C), 
as well as the lensing galaxy,  are identified in the SDSS imaging pipeline. 
The lensing galaxy is located in the field of a galaxy cluster. The red box represents the  
galaxy SDSS J082728.62+223241.0, which has an SDSS spectrum with z=0.3346. North 
is up and East is to the left in this figure.}
\label{fig:sdss_cutout}
\end{figure}

\begin{figure}
\plotone{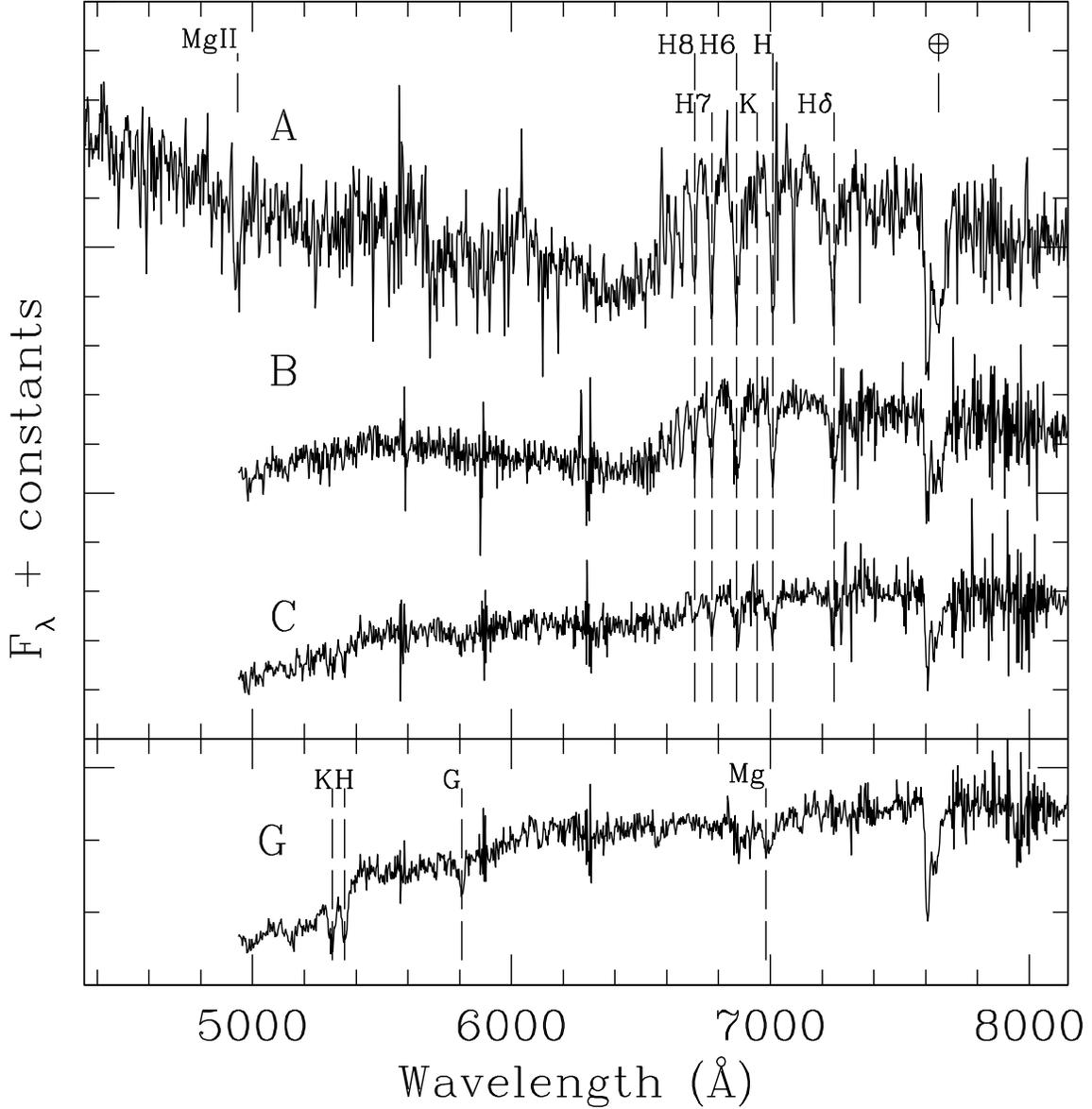}
\caption{Spectra of the lens images (A, B, and C) and the lensing galaxy (G). Wavelength is 
given in the observed frame. The spectra 
of the lens images A, B, and C have the same redshift of 0.766. The lensing galaxy has z = 0.349 
as measured from the CaII K \& H, G-band, and Mg lines. Spectra B, C, and G were obtained using the 
Subaru FOCAS instrument, while the spectrum of A is from the APO 3.5m DIS spectrograph. The inconsistency  
of the continua of A and B is due to slit losses from atmospheric refraction.}
\label{fig:spectrum}
\end{figure}

\begin{figure}
\plotone{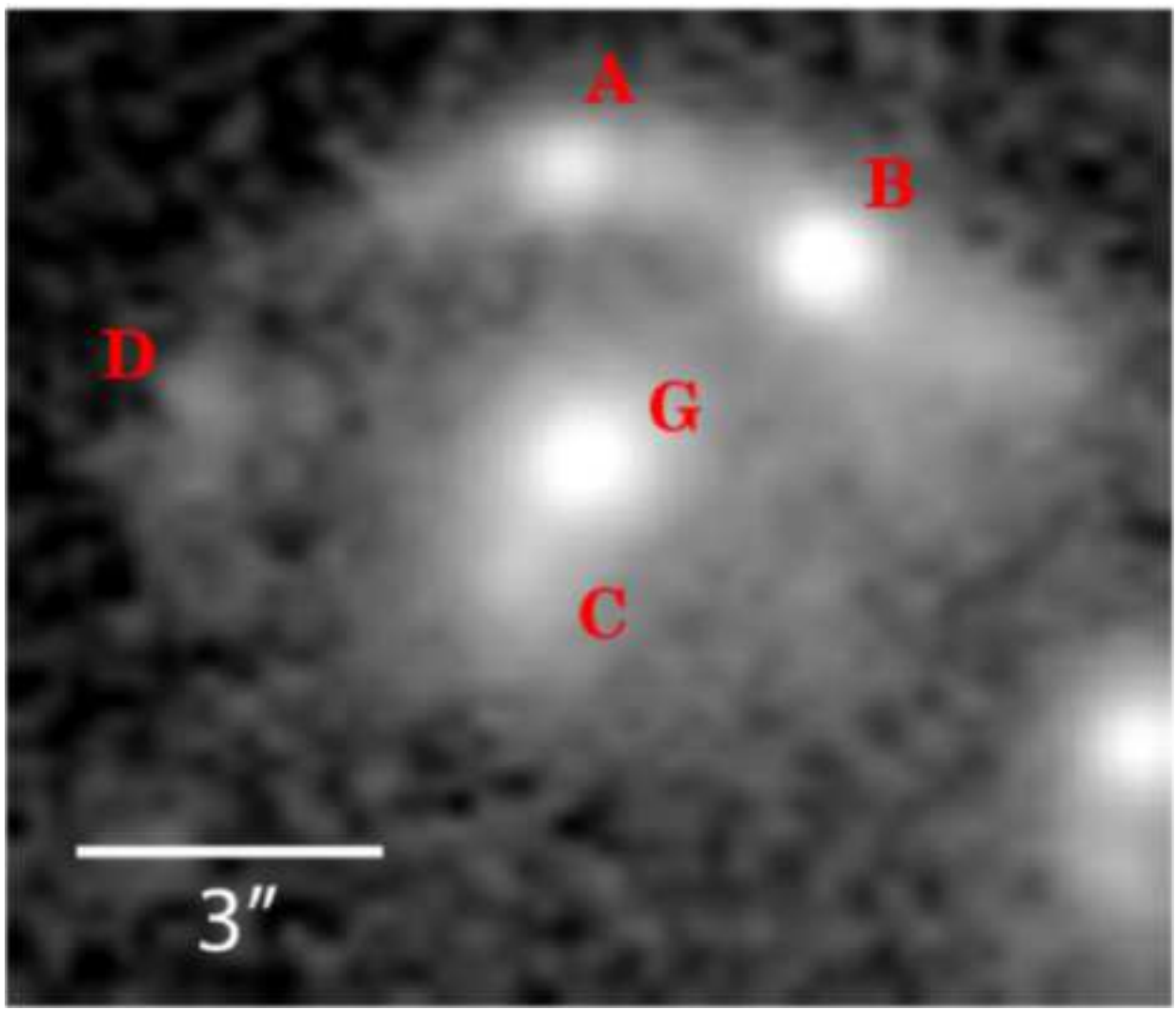}
\caption{Subaru $V$-band image. The brightness peak positions of objects A, B, C, D, and G are measured 
using SExtractor. North is up and East is to the left.}
\label{fig:Subaru_V}
\end{figure}

\begin{figure}
\plottwo{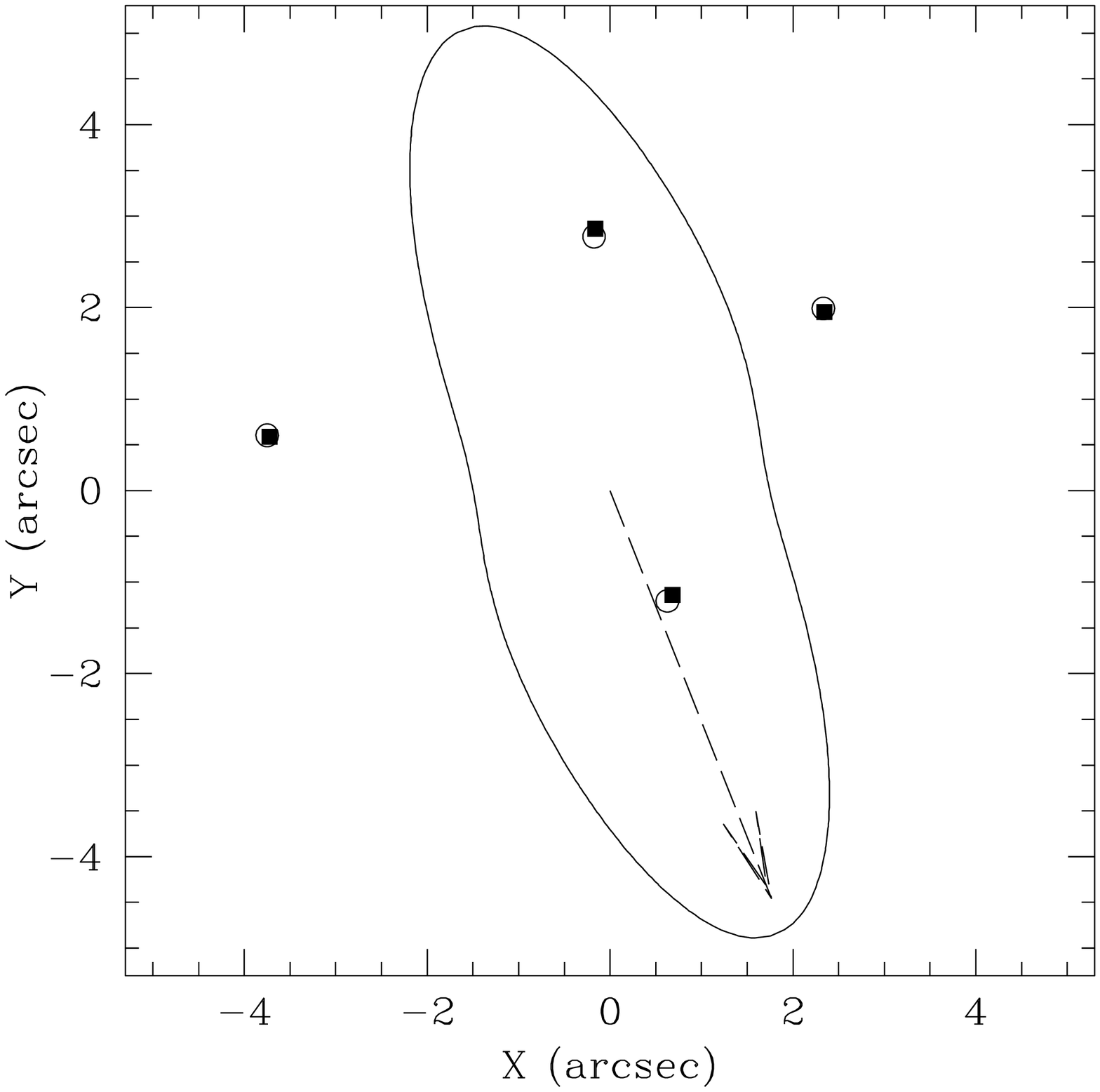}{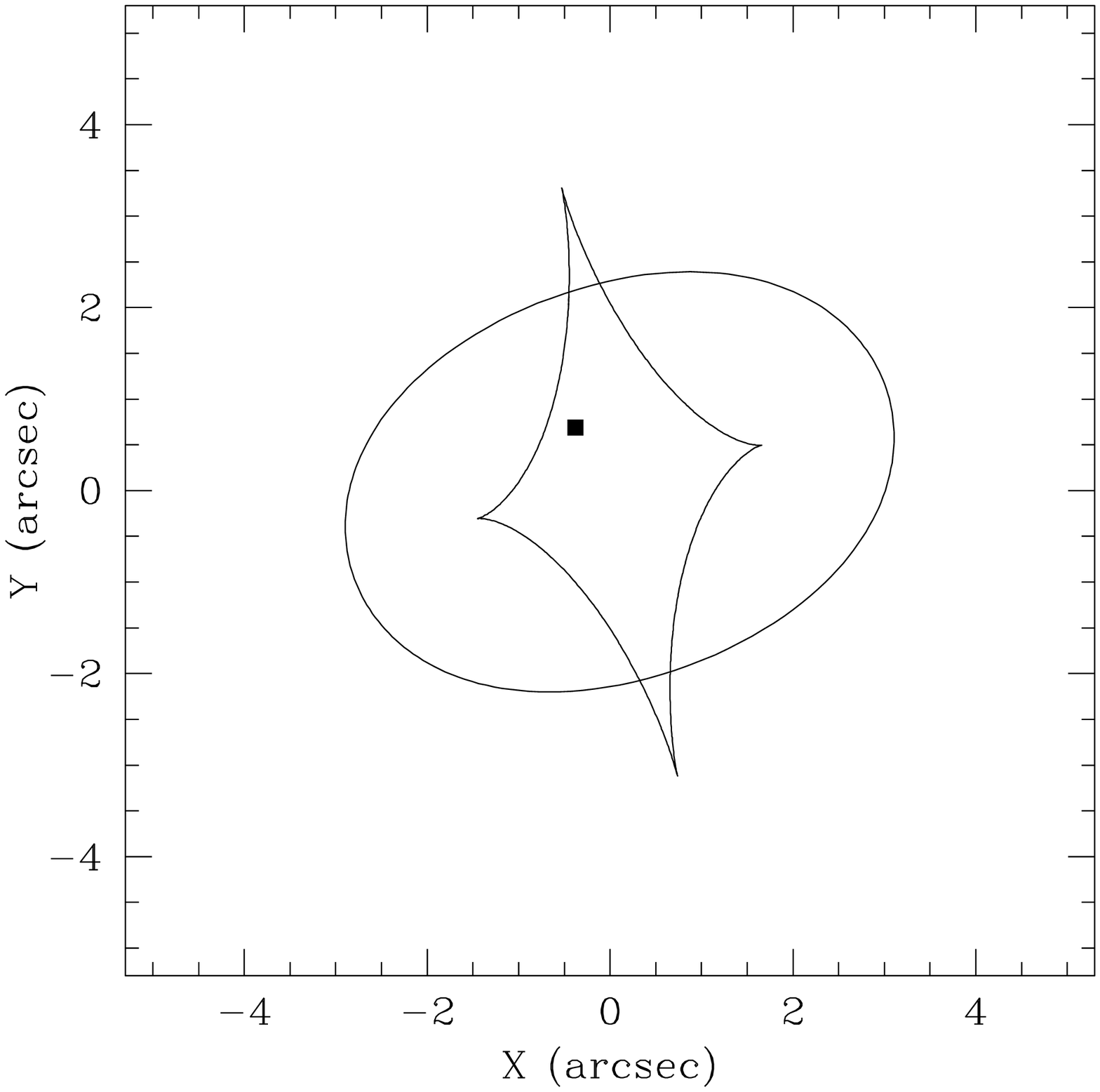}
\caption{Results of fitting a singular isothermal ellipsoid with 
external shear to the positions of the four images. In the image plane {\it (left)}, 
the observed positions of lens images are shown as filled rectangles, 
while the fitted positions 
are given as open circles. The difference of the positions is $\sim$ 0\farcs2. The critical 
curve of the fitted lens model {\it (solid line)} shows that the external shear plays 
an important role and the direction of the quadrupole term is 
toward the brightest cluster galaxy {\it (dashed arrow)}. 
In the source plane {\it (right)}, the derived source position 
{\it (filled rectangle)} is inside both the outer pseudo-caustic and 
the astroid caustic.}
\label{fig:lens_model}
\end{figure}

\begin{figure}
\plotone{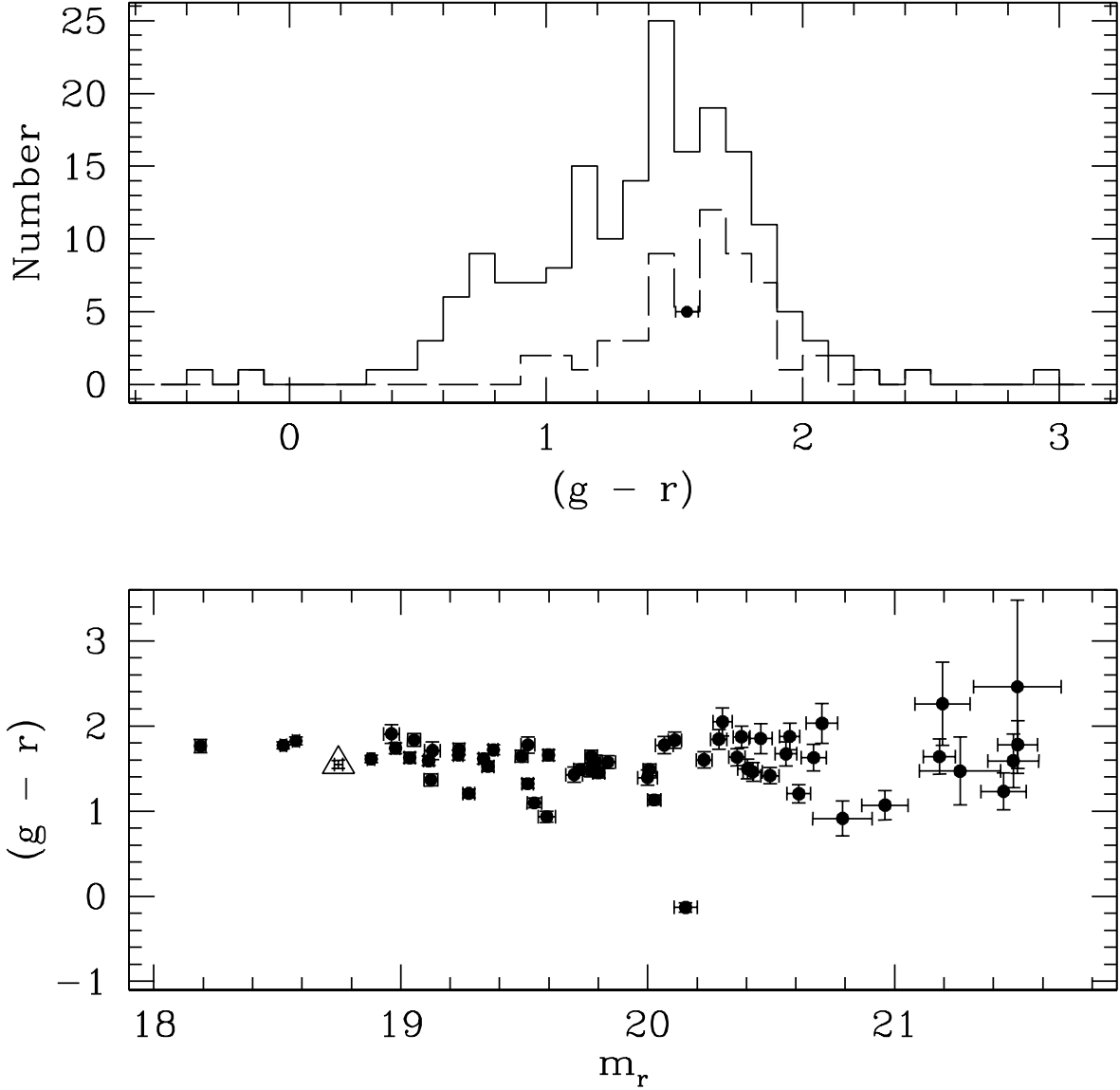}
\caption{Color distribution and color-magnitude diagram of galaxies within 3\arcmin\ of the lensing galaxy 
from SDSS imaging. 
{\it (top)} The solid line shows the distribution of galaxies brighter than $r = 21.5$ within 
3\arcmin, while 
the dashed line represents the color distribution of ellipticals only. 
The color of the lensing galaxy 
is shown as a point with 1$\sigma$ error. {\it (bottom)} The color-magnitude 
diagram of ellipticals shows 
the red sequence of a galaxy cluster. The lensing galaxy {\it (open triangle)}
is not the brightest elliptical in this cluster. All magnitudes and colors are corrected for Galactic 
extinction based on \citet{schlegel98}.}
\label{fig:cluster}
\end{figure}

\end{document}